\documentclass{emulateapj}

\usepackage{rotating}

\begin{document}

\shorttitle{Spectroscopy of Andromeda~X}
\shortauthors{Kalirai et al.}

\title{The SPLASH Survey: A Spectroscopic Analysis of the Metal-Poor, \\ Low-Luminosity 
M31 dSph Satellite Andromeda~X$^{1,2}$}


\author{
Jason S.\ Kalirai\altaffilmark{3,4},
Daniel B.\ Zucker\altaffilmark{5,6,7,8},
Puragra Guhathakurta\altaffilmark{4},
Marla Geha\altaffilmark{9}, \\
Alexei Y.\ Kniazev\altaffilmark{5,10,11}, 
David Mart\'{i}nez-Delgado\altaffilmark{5,12,13},
Eric F. Bell\altaffilmark{5}, \\
Eva K.\ Grebel\altaffilmark{14}, and
Karoline M.\ Gilbert\altaffilmark{15}
}
\altaffiltext{1}{Data presented herein were obtained at the W.\ M.\ Keck
Observatory, which is operated as a scientific partnership among the
California Institute of Technology, the University of California, and the
National Aeronautics and Space Administration.  The Observatory was made
possible by the generous financial support of the W.\ M.\ Keck Foundation.}
\altaffiltext{2}{Based on observations made with the Nordic Optical Telescope, 
operated on the island of La Palma jointly by Denmark, Finland, Iceland, Norway, 
and Sweden, in the Spanish Observatorio del Roque de los Muchachos of the Instituto 
de Astrof\'{i}sica de Canarias; these observations were funded by the 
Optical Infrared Coordination Network (OPTICON), a major international 
collaboration supported by the Research Infrastructures Programme of the 
European Commission's Sixth Framework Programme.}
\altaffiltext{3}{Space Telescope Science Institute, 3700 San Martin Drive, 
Baltimore MD, 21218; jkalirai@stsci.edu}
\altaffiltext{4}{University of California Observatories/Lick Observatory, 
University of California at Santa Cruz, 1156 High Street, Santa Cruz CA, 
95064, USA; raja@ucolick.org}
\altaffiltext{5}{Max-Planck-Institut f\"{u}r Astronomie, K\"{o}nigstuhl 17, 
D-69117 Heidelberg, Germany; bell@mpia.de}
\altaffiltext{6}{Institute of Astronomy, University of Cambridge, Madingley Road, 
Cambridge, CB3~0HA, United Kingdom}
\altaffiltext{7}{Department of Physics, Macquarie University, 
North Ryde, NSW 2109, Australia; zucker@science.mq.edu.au}
\altaffiltext{8}{Anglo-Australian Observatory, PO Box 296, Epping, 
NSW 1710, Australia}
\altaffiltext{9}{Astronomy Department, Yale University, New Haven CT, 
06510; marla.geha@yale.edu}
\altaffiltext{10}{South African Astronomical Observatory, PO Box 9, 7935 Observatory, 
Cape Town, South Africa; akniazev@saao.ac.za}
\altaffiltext{11}{South African Large Telescope Foundation, PO Box 9, 7935 Observatory, 
Cape Town, South Africa}
\altaffiltext{12}{Instituto de Astrof\'{i}sica de Canarias, 
La Laguna, Spain; ddelgado@iac.es}
\altaffiltext{13}{Ram\'{o}n y Cajal Fellow}
\altaffiltext{14}{Astronomisches Rechen-Institut, Zentrum f\"ur Astronomie der 
Universit\"at  Heidelberg, M\"onchhofstr.\ 12--14, D-69120 Heidelberg, Germany; 
grebel@ari.uni-heidelberg.de}
\altaffiltext{15}{Department of Astronomy, Box 351580, University of Washingon, 
Seattle WA, 98195; kmgilber@u.washington.edu}


\begin{abstract}

\noindent Andromeda~X (And~X) is a newly discovered low-luminosity M31 dwarf spheroidal 
galaxy (dSph) found by Zucker et~al.\ (2007) in the Sloan Digital Sky Survey (SDSS -- York 
et~al.\ 2000).   In this paper, we present the first spectroscopic study of individual red 
giant branch stars in And~X, as a part of the SPLASH Survey (Spectroscopic and Photometric 
Landscape of Andromeda's Stellar Halo).  Using the Keck~II telescope and multiobject 
DEIMOS spectrograph, we target two spectroscopic masks over the face of 
the galaxy and measure radial velocities for $\sim$100 stars with a median 
accuracy of $\sigma_v$ $\sim$ 3~km~s$^{-1}$.  The velocity histogram for 
this field confirms three populations of stars along the sight line: foreground Milky 
Way dwarfs at small negative velocities, M31 halo red giants over a broad 
range of velocities, and a very cold velocity ``spike'' consisting of 22 stars 
belonging to And~X with $v_{\rm rad} = -163.8 \pm 1.2$~km~s$^{-1}$.  By carefully 
considering both the random and systematic velocity errors of these stars (e.g., 
through duplicate star measurements), we derive an intrinsic velocity dispersion 
of just $\sigma_v = 3.9 \pm 1.2$~km~s$^{-1}$ for And~X, which for its size, 
implies a minimum mass-to-light ratio of $M/L_V$ = $37^{+26}_{-19}$ assuming 
the mass traces the light.  Based on the clean sample of member stars, we 
measure the median metallicity of And~X to be [Fe/H] = $-$1.93 $\pm$ 0.11, 
with a slight radial metallicity gradient.  The dispersion in metallicity is 
large, $\sigma$([Fe/H]$_{\rm phot}$) = 0.48, possibly hinting that the galaxy 
retained much of its chemical enrichment products.  \\

\noindent And~X has a total integrated luminosity ($M_V$ = $-$8.1 $\pm$ 0.5) that 
straddles the classical Local Group dSphs and the new SDSS ultra-low luminosity 
galaxies.  The galaxy is among the most metal-poor dSphs known, especially relative 
to those with $M_V < -$8, and has the second lowest intrinsic velocity dispersion 
of the entire sample.  Our results suggest that And~X is less massive by a factor 
of four when compared to Milky Way dSphs of comparable luminosity (e.g., Draco 
and Ursa Minor).  We discuss the potential for better understanding the formation 
and evolution mechanisms for M31's system of dSphs through (current) 
kinematic and chemical abundance studies, especially in relation to the Milky 
Way sample.


\end{abstract}

\keywords{dark matter -- galaxies: abundances -- galaxies: dwarf -- galaxies: 
individual (And~X) -- techniques: spectroscopic}


\section{Introduction} \label{intro}

A census of dwarf galaxies in the neighborhoods of large systems 
such as the Milky Way and Andromeda (M31) can yield important constraints on 
the processes that shaped massive galaxies.  Current 
predictions from $\Lambda$CDM simulations suggest that at least an order of 
magnitude more satellites should be present around these systems than has 
been observed (e.g., Klypin et~al.\ 1999; Moore et~al.\ 1999).  Several 
explanations for this ``missing satellite problem'' have been suggested, both 
on the theoretical and observational fronts.  For example, many of the 
predicted subhalos may be entirely ``dark'', and therefore may not have 
experienced any star formation (e.g., Somerville 2002), and/or our census of 
these low-luminosity galaxies may be highly incomplete.

Recent wide field surveys in both the Milky Way and M31 have confirmed the 
latter possibility.  In our Galaxy, the number of dwarf spheroidal galaxies 
(dSphs) known has doubled in the past few years, thanks in large part to the Sloan 
Digital Sky Survey (SDSS -- York et al. 2000; e.g., Willman et~al.\ 2005; Zucker et~al.\ 2006a, 
Belokurov et~al.\ 2006; Zucker et~al.\ 2006b; Belokurov et~al.\ 2007; Walsh, 
Jerjen, \& Willman 2007).  Similarly, the last three years has seen the number of 
dSphs in M31 grow by more than a factor of two 
\citep{zucker04b,zucker07,martin06,majewski07,ibata07,mcconnachie08}.  
Yet, considering the limited regions of the Milky Way and M31 that have been 
surveyed for these lower luminosity galaxies, many more such galaxies should be 
discovered in the coming years (e.g., with Pan-STARRS, the Southern Hemisphere 
Stromlo Missing Satellites Survey on the SkyMapper Telescope -- Keller et~al.\ 
2007, and the Large Synoptic Survey Telescope, see Tollerud et~al.\ 2008 for 
expected number counts).

None of the newly discovered Milky Way dSphs in the SDSS are directly seen in the 
initial images, but rather they are detected as resolved stellar overdensities.  Deeper 
follow up images have revealed that some of these galaxies are distinct in their properties 
as compared to the existing sample; they are more irregular and have very low 
surface brightnesses ($\mu_V >$ 27 mag~arcsec$^{\rm -2}$).  The faintest objects, 
such as Willman~1 and Segue~1, have an integrated luminosity less than a {\it single 
bright red giant}, and in an $M_V$ vs half-light radius plane, appear to bridge 
globular clusters and dwarf galaxies \citep{belokurov07}.  Several investigations 
are now underway to understand the detailed nature of these systems (e.g., mean 
abundance, abundance spread, velocities, and mass-to-light ratios) through follow 
up spectroscopic observations (e.g., Simon \& Geha 2007; Koch et~al.\ 2008a; Norris 
et~al.\ 2008; Geha et~al.\ 2009).  

In distinction to the Milky Way satellites, there have been limited spectroscopic 
observations of individual stars in M31 dSphs, with most studies only confirming a 
handful of members \citep{cote99,guh00,chapman05,majewski07}.  Similar studies of 
M31 {\it field stars} in the SPLASH Survey have recently revealed that the metal-rich 
inner spheroid of M31 dominates the metal-poor outer halo out to a much larger 
radius (20 -- 30~kpc) than in the Milky Way \citep{guh05,kalirai06a}.  This 
disparity likely reflects a more violent accretion history in M31 as compared to 
our Galaxy \citep{hammer07}, the roots of which may lie in the 
properties of the accreted (and possibly unaccreted) satellite systems of the 
two galaxies.  In fact, \cite{mcconnachie06} show that, at a fixed luminosity, 
the scale radii of the M31 dSph galaxies is a factor of two larger than the 
canonical Milky Way satellites.  If the dark matter masses of these galaxies are 
the same {\it and} the mass traces the light, this difference should be reflected 
in the velocity dispersions of the galaxies with the M31 sample being colder.  
However, alternate assumptions on the spatial distribution of stars relative to the 
dark matter can lead to different conclusions, as discussed in \S\,\ref{discussion}.

To address some of these questions, our team has now begun a systematic survey of M31 
dSphs with the Keck~II telescope and DEIMOS multi-object spectrograph.  At the 
distance of M31, the apparent angular size of a typical dSph is well suited for 
this instrument as we can easily target several hundred potential red giants 
simultaneously.  To date, six of the M31 dSphs have been targeted, including 
And~XIV which has been presented in \cite{majewski07} and And I, II, and III which 
will be presented in \cite{kalirai09}.  In this paper, we discuss the {\it first} 
spectroscopic study of individual red giants in the newly discovered, low-luminosity 
M31 dSph, And~X \citep{zucker07}.  Our target selection, observational design, 
and data reduction are presented in the next section, followed by an analysis of 
the And~X velocity histogram to establish both the membership of individual stars 
and the mean velocity of the galaxy (\S\,3).  We resolve the intrinsic central 
velocity dispersion of And~X in \S\,\ref{velocitydispersion} and calculate the 
mass-to-light ratio of the galaxy.  Chemical abundances of the member stars in 
And~X are calculated both photometrically and spectroscopically in \S\,4.  The 
results are discussed and summarized in \S\,5.


\begin{figure}
\begin{center}
\leavevmode
\includegraphics[width=8.5cm]{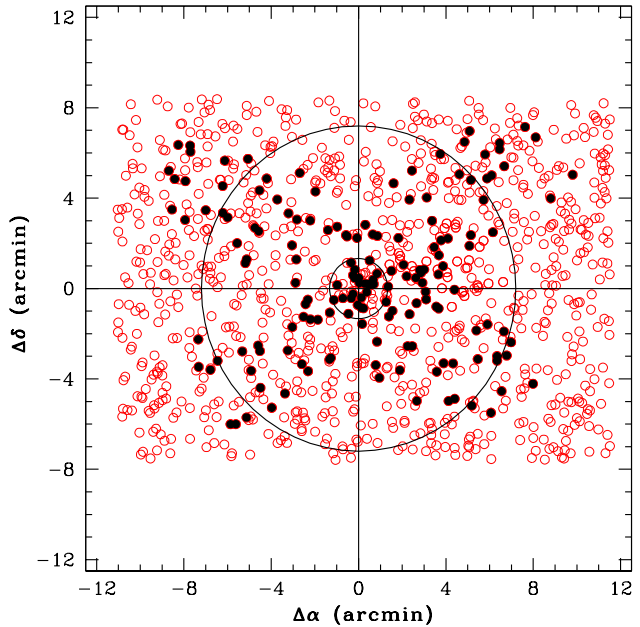}
\end{center}
\caption{The selection of Keck/DEIMOS spectroscopic targets (filled 
circles) from the photometric catalog (open circles) of And~X.  The 
straight lines intersect at the center of And~X ($\alpha_{\rm J2000}$ = 
01$^{\rm h}$06$^{\rm m}$39.20$^{\rm s}$, $\delta_{\rm J2000}$ = 
44$^{\rm \circ}$48$'$44.6$''$) and the two circles 
denote the core radius (1$\farcm$33) and tidal radius (7$\farcm$2) of 
the galaxy \citep{zucker07}.  This ``cross'' pattern reflects the 
$\pm$45~degree position angle of the two spectroscopic masks.  This 
arrangement ensures the targeting of many stars near the center of 
And~X as well as along spokes that extend to beyond the tidal radius 
of the galaxy.}
\label{fig:d10starcountmap}
\end{figure}



\begin{figure*}
\begin{center}
\leavevmode
\includegraphics[width=12cm,angle=270]{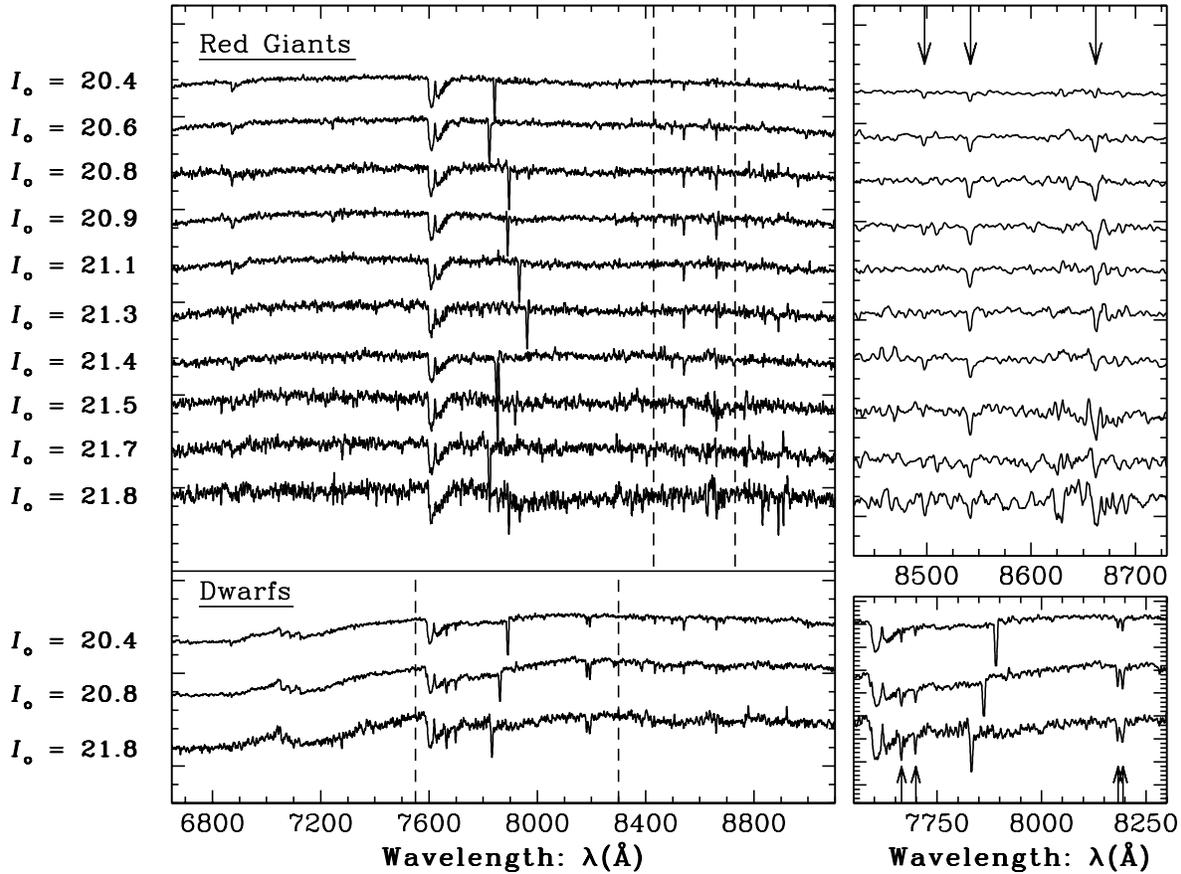}
\end{center}
\caption{The spectra of red giants (top) and three foreground Milky Way dwarfs 
(bottom) in our And~X data set, smoothed by a 10~pixel boxcar function and 
redshifted to zero velocity.  The B-band and A-band atmospheric absorption features 
can be seen in the spectra at 6900~${\rm \AA}$ and 7600~${\rm \AA}$.  A dip close 
to the center of each spectrum is caused by the CCD gap.  The stars presented here 
have been chosen to reflect the full magnitude range over which we measure radial 
velocities.  A closer look at some of the key spectral features for these stars is 
presented in the two panels on the right.  For the giants (top), we display the region 
of the Ca\,{\sc ii} triplet, with an arrow at the top indicating the position of 
each line.  For the dwarfs (bottom), we focus on the pressure sensitive Na\,{\sc i} doublet 
(8190~${\rm \AA}$) and K\,{\sc i} lines (7665 and 7699~${\rm \AA}$) that are used to 
eliminate foreground contamination (see \S\,\ref{velocity}).}
\label{fig:d10spectra}
\end{figure*}


\section{The Data}\label{data}

\subsection{Observations}\label{observations}

And~X was first discovered as a stellar overdensity on an SDSS imaging scan of 
the major axis of M31 (extending 18$^{\rm o}~\times$ 2.5$^{\rm o}$), observed on 
5 October 2002 \citep{zucker04a}.  The overdensity was only visible after 
selecting objects with magnitudes and colors consistent with the red 
giant branch in a metal-poor population, at the distance of M31.  Spatially, 
the overdensity lies along the projection of the N-E major axis at a distance of 
$\sim$80~kpc from M31's center.  These 
initial observations were followed up with subsequent imaging from the William 
Herschel Telescope and the Nordic Optical Telescope in late 2004.  The resulting 
images confirmed the presence of a low surface brightness, spatially coherent 
group of stars with a color-magnitude diagram (CMD) indicating a blue red giant branch 
(i.e., presumably metal-poor stars).  The integrated $V$-band luminosity of the galaxy 
is $M_V$ = $-$8.1 $\pm$ 0.5.  Further details on these photometric observations are 
presented in \cite{zucker07}, although we note here that our default photometric 
catalogue is taken to be the William Herschel Telescope data as the calibration is 
superior to the Nordic Optical Telescope data.

We used these photometric data as input for our multiobject spectroscopic 
observations with Keck~II.  The image of And~X from the Nordic Optical 
Telescope shows the galaxy to have an extent of a few arcminutes.  The 
estimated tidal radius by \cite{zucker07} is $\sim$7~arcminutes, much 
smaller than the long-axis of the DEIMOS spectrograph, which extends 
$\sim$16~arcminutes (see Figure~\ref{fig:d10starcountmap}).  Therefore, we 
could easily observe the full extent of the galaxy with a 
single pointing.  We selected targets based on their location in the 
CMD.  The tip of the red giant branch in And~X is 
located at $I_{\rm 0}$ = 20.25 ($d$ = 701~kpc), and so our highest priority 
objects are those with 20.0 $< I_{\rm 0} <$ 22.0.  Lower priorities are given 
to fainter stars.  


\begin{table*}
\begin{center}
\caption{}
\begin{tabular}{lcccccccc}
\hline
\hline
\multicolumn{1}{c}{Date} & \multicolumn{1}{c}{Mask} & \multicolumn{1}{c}{$\alpha_{J2000}$} & 
\multicolumn{1}{c}{$\delta_{J2000}$} & \multicolumn{1}{c}{Pos. Angle ($^{\rm o}$)} & 
\multicolumn{1}{c}{Exp. Time (s)} & \multicolumn{1}{c}{Seeing} & 
\multicolumn{1}{c}{Airmass} & \multicolumn{1}{c}{No. of Targets$^*$}\\ 

\hline
5 Sept 2005 & d10$\_$1 & 01:06:39.20 & 44:48:44.6  & 45    & 3$\times$1200 &  0$\farcs$65 -- 0$\farcs$77 & 1.26 -- 1.41 & 85 \\
5 Sept 2005 & d10$\_$2 & 01:06:39.20 & 44:48:44.6  & $-$45 & 3$\times$1200 &  0$\farcs$51 -- 0$\farcs$60 & 1.14 -- 1.21 &
94 \\

\hline
\end{tabular}
\tablenotetext{$^*$}{14 objects were targeted on both masks.}
\label{table1}
\end{center}
\end{table*}


Unfortunately, the stellar density over the And~X field of view is not 
high enough to utilize the full capabilities of DEIMOS, an instrument that 
could easily target $\gtrsim$200 stars in a single pointing.  Although we can increase the 
number of And~X targets by pushing to fainter limits along the red giant 
branch, our experience has shown that it is difficult to measure accurate 
radial velocities from such lower signal-to-noise data unless the integration 
times are made much longer.  We therefore, generate a second spectroscopic 
mask at roughly the same position, but offset in position angle by 
90~degrees (shown in Figure~\ref{fig:d10starcountmap}).  This offers us two 
advantages.  First, we can target {\it mostly} different stars in the two 
pointings to increase the number of reliable measurements near the tip of 
the red giant branch in And~X.  Second, we can obtain duplicate spectra for a few 
stars in order to constrain the uncertainties in our radial velocity 
measurements based on independent measurements.  In total, we targeted 
85 objects on the first mask (d10$\_$1) and 94 objects on the second 
mask (d10$\_$2), 14 of which were placed on both masks.

We observed each of these two spectroscopic masks for 1~hour on 5 September 
2005 (see Table~1).  The individual exposure times were set at 1,200 seconds.  
The observing conditions were photometric, the airmass was low at 1.2 -- 1.4, 
and the seeing was exceptional at 0$\farcs$51 -- 0$\farcs$65 for most of the 
observations, and at worst 0$\farcs$77 for one exposure.  The observations 
were obtained using our standard setup, as described in several previous papers 
(e.g., Gilbert et~al.\ 2006).  This utilizes the 1200~lines~mm$^{-1}$ grating at 
a central wavelength of $\rm7800\AA$, with a wavelength range of approximately 
$\approx6400$ -- $\rm9100\AA$ for each star.

\subsection{Data Reduction} \label{datareduction}

The two DEIMOS multi-slit masks for And~X were reduced using the 
{\tt spec2d} software pipeline developed by the DEEP2 team at the 
University of California-Berkeley (see Faber et~al.\ 2007 for 
details).  To summarize the key steps, the flat-field exposures are 
first used to rectify the curved raw spectra into rectangular arrays by
applying small shifts and interpolating in the spatial direction.
A one-dimensional slit function correction and two-dimensional
flat-field and fringing correction are applied to each slitlet.
Using the DEIMOS optical model as a starting point, a
two-dimensional wavelength solution is determined from the arc
lamp exposures with residuals of order $\rm0.01\mbox{\AA}$.  Each
slitlet is then sky-subtracted exposure by exposure using a
B-spline model for the sky.  The individual exposures of the
slitlet are averaged with cosmic-ray rejection and
inverse-variance weighting.  Finally one dimensional spectra are
extracted for all science targets using the optimal scheme of
\cite{hor86} and are rebinned into logarithmic wavelength bins
with 14~km~s$^{-1}$ per pixel.  Example spectra for several stars 
are shown in Figure~\ref{fig:d10spectra}.

We measure radial velocities for all objects by cross-correlating 
the spectra with high signal-to-noise stellar templates.  The methods used 
for this analysis are described in detail in \cite{simon07}.  
Specifically, we note that a correction is made to offset any wavelength calibration 
error that results from imprecise astrometry.  This is performed by cross-correlating 
each spectrum with several bright, hot standard stars.  These stars exhibit a 
nearly featureless spectrum, but with a clean detection of the telluric lines in 
the A-band absorption feature.  The resulting cross-correlation leads to an estimate of 
how much the object was offset from the center of the slit and therefore 
provides a means to correct any resulting wavelength (and therefore velocity) 
error (see also Sohn et~al.\ 2006 for further details).  

Of the 179 spectra obtained, this cross-correlation yielded an accurate 
velocity for just over half of the stars (98 in total).  We manually verified 
each of the template fits and only kept those which showed at least two 
definite spectral features in common.  In fact, for most objects multiple 
absorption features were matched over the 3000~${\rm \AA}$ wavelength range.  
Additionally, we kept 15 objects in the group of 98 whose spectra showed at 
least one definite match to the template {\it and} a second marginal match 
(see discussion in Guhathakurta et~al.\ 2006; Kalirai et al.\ 2006b).  The 
objects for which this failed typically have poor signal-to-noise spectra 
and are predominantly the fainter targets ($I_{\rm 0} >$ 22) in our study.  
Finally, we applied a heliocentric correction to each star's velocity.

Of the original 14 duplicate stars, 7 have measured velocities on both masks, 
one of which is in the marginal category discussed above.  The standard 
deviation in the repeat velocity measurements between the two independent 
measurements for these objects is 4.1~km~s$^{-1}$, and the mean error in 
velocity among the 14 measurements is 4.0~km~s$^{-1}$.  The 
differences are not systematically correlated with any one set of observations.  For 
our final velocity catalog, we calculate the weighted mean velocity of each 
of the duplicate pairs and therefore have 91 individual stars.


\section{Kinematics}\label{kinematics}

\subsection{Radial Velocity Accuracy}\label{accuracy}

Without careful treatment, radial velocities derived from relatively low 
resolution spectra can have individual velocity uncertainties of at least 
several km~s$^{-1}$.  In fact, not applying the A-band absorption 
correction discussed earlier would alone translate to an uncertainty of 
$\sim$3~km~s$^{-1}$ for these data.  This has been directly tested by 
Simon \& Geha (2007) by comparing the measured velocities of duplicate stars before 
and after the correction.  As typical low-luminosity Milky 
Way dSphs are kinematically cold, with velocity dispersions of $\sigma_v \lesssim$ 
10~km~s$^{-1}$, an accurate measurement of the radial velocities, and an understanding 
of the velocity errors, are crucial to probe the internal kinematics of 
these galaxies.  


\begin{figure*}
\begin{center}
\leavevmode
\includegraphics[width=12cm,angle=270]{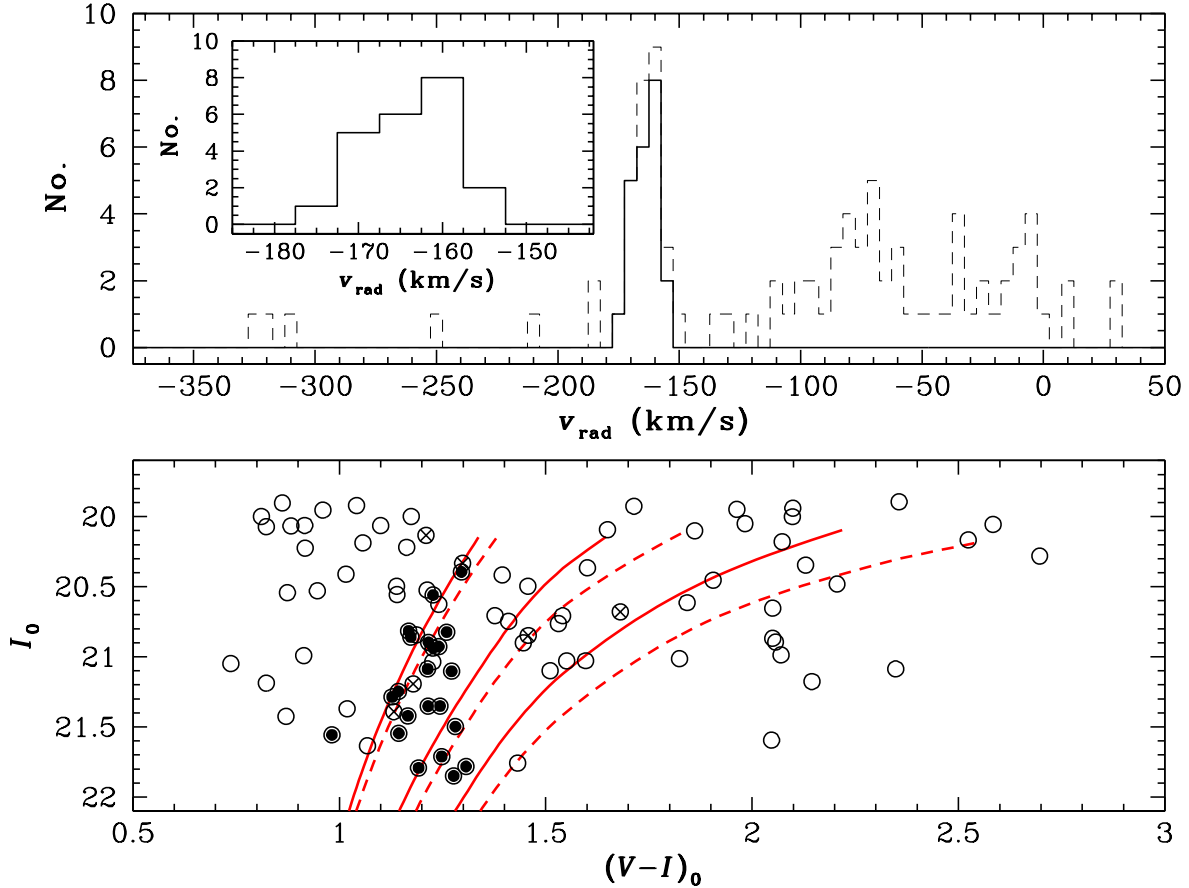}
\end{center}
\caption{{\it Top --} The velocity histogram of all objects measured 
along the line of sight (dashed line) clearly indicates a kinematically 
cold population of stars with $v_{\rm rad}$ = $-$163.8 $\pm$ 1.2~km~s$^{-1}$, 
and $\sigma_v$ = 3.9 $\pm$ 1.2~km~s$^{-1}$.  This velocity ``spike'' represents 
And~X.  At smaller negative velocities, a large number of Milky Way dwarfs are 
present in the data set.  These objects, as well as candidate M31 halo giants 
at larger negative velocities, are removed as discussed in \S\,\ref{velocity} 
to yield the final And~X velocity histogram (solid line).  A finer scale, 
centered on the dSph's radial velocity, is shown in the inset.  
{\it Bottom --} And~X velocity members (filled circles) are 
found to lie in a tight sequence consistent with a metal-poor red giant 
branch population.  The solid red curves indicate [$\alpha$/Fe] = 0.0 
isochrones (dashed are for [$\alpha$/Fe] = $+$0.3) with metallicities of 
[Fe/H] = $-$2.31 (left), $-$1.31 (middle), and $-$0.83 (right) and ages of 
12~Gyr from \cite{vandenberg06}.  The open circles designate the non-member stars along 
the line of sight and the six crosses with open circles mark the objects 
that we further rejected from the initial sample as discussed in 
\S\,\ref{velocity}.  Stars with $I_{\rm 0}$ $\gtrsim$ 21.5 are affected by larger 
photometric errors, $\sigma_{(V-I)}$ $\sim$ 0.2, as discussed in the text.}
\label{fig:d10cmdvel2}
\end{figure*}


We measure the errors in our velocity measurements using the method 
discussed by \cite{simon07}.  First, we generate 500 spectra with artificial 
noise at each pixel scaled by the observed variance in that pixel, {\it for 
each star}.  This Monte Carlo leads to an error bar which is defined as the 
square root of the variance recovered from re-fitting these 500 spectra.  The 
error estimates are next compared to the difference in velocity of the duplicate 
measurements to establish the contribution from any additional error sources 
that are not accounted for in the Monte Carlo method.  \cite{simon07} 
estimated this ``floor'' to be $\epsilon$ = 2.2~km~s$^{-1}$ in their data set. This 
implies that the observed difference in velocity measured for duplicate stars 
(taken to be the true error in the measurements), is reproduced by adding the 
quadrature sum of the Monte Carlo error estimates with a 2.2~km~s$^{-1}$ 
additional error bar.  For our duplicates (including stars observed in And~I, 
II, and III; will be presented Kalirai et~al.\ 2009), we measure a value of $\epsilon$ $\lesssim$ 
3~km~s$^{-1}$, very similar to \cite{simon07}.  As their measurement is based 
on a factor of three larger sample than ours, we fold in a 2.2~km~s$^{-1}$ 
error to the measured Monte Carlo values.  

As discussed in \S\,2.3 of \cite{simon07}, a comparison of the velocities and 
velocity dispersions of kinematically cold populations with measurements and errors 
as described above yields excellent agreement with independent high-resolution analysis.  
Specifically, for a similar number of stars, the measured velocity dispersion of the 
globular cluster NGC~2419 using Keck/DEIMOS in this setup is identical to HIRES measurements 
discussed in \cite{simon07}.  Similarly, the comparison of a half dozen Keck/DEIMOS red giant 
spectroscopic measurements to high-resolution analysis of the same stars in UMa~I 
by \cite{kleyna05} indicates differences that are smaller than the 1$\sigma$ uncertainty.  
These results illustrate that the method above can be used to recover velocity 
dispersions as low as 2.3~km~s$^{-1}$. 

\subsection{Membership and the Mean Velocity of And~X}\label{velocity}

\cite{zucker07} estimate the distance modulus of And~X to be 
($m - M$)$_{\rm 0} \sim$ 24.23 from the tip of the red giant branch.  
This places And~X $\sim$80~kpc in front of M31.  Our line of sight towards 
And~X will therefore first intersect the disk and halo of the Milky Way, 
then And~X, and finally M31's stellar halo.  The method that we have used 
to select targets is based on the photometric properties of the stars, and 
therefore our resulting radial velocity catalog contains a mix of stars 
belonging to each of these three stellar populations (e.g., a nearby Milky 
Way dwarf can mimic an M31 red giant in the CMD).  Fortunately, the radial 
velocity signatures of each of these groups of stars is expected to be 
different.  

The radial velocity histogram for the 91 stars in our final catalog is 
presented as a dashed line in Figure~\ref{fig:d10cmdvel2} (top).  The 
most salient feature in the diagram is the kinematically cold ``spike'' 
in the center.  This velocity peak represents And~X and therefore 
kinematically confirms the presence of a new dwarf satellite galaxy in the \cite{zucker07} 
imaging study.  In addition to this population, we see an asymmetric 
velocity distribution around And~X.  Many more stars are found at smaller 
negative velocities, suggesting that a significant number of Milky Way 
dwarfs are in our sample.  This is expected for several reasons.  First, And~X 
is a low luminosity galaxy and only occupies a small fraction of each 
of our DEIMOS masks, the rest of which is filled predominantly with 
non-member stars (see Figure~\ref{fig:d10starcountmap}).  Second, the input 
photometry that we used was not pre-screened to eliminate possible 
dwarfs based on their $DDO51$ photometry, as was done by members of our 
group in studies of And~XIV (Majewski et~al.\ 2007) and M31's stellar halo 
(Guhathakurta et~al.\ 2005; Kalirai et~al.\ 2006a).

\cite{gilbert06} show that the bulk of the Milky Way dwarfs along M31's 
line of sight have a small negative radial velocity.  Still, these stars 
can overlap the much (dynamically) hotter spheroid or halo of M31, which 
has a mean velocity of $v_{\rm rad}$ = $-$300~km~s$^{-1}$.  \cite{gilbert06} go 
on to develop a method based on several photometric and spectroscopic 
diagnostics that efficiently cleans out the foreground dwarf star contamination 
from M31 halo giants.  Specifically, this method combines information from the 
velocity histogram, color-magnitude diagram, metallicity measurements using 
two approaches, and also incorporates the strengths of several spectral features 
to yield a clean sample of giants.   For example, the measurement 
of the Na\,{\sc i} doublet at 8190~${\rm \AA}$ and the K\,{\sc i} absorption 
lines at 7665 and 7699~${\rm \AA}$ help yield excellent discrimination between 
cool dwarfs and giants as these lines are pressure sensitive, and therefore strong 
in dwarfs and absent in giants.  These absorption lines can be easily seen in the 
raw spectra of the cooler dwarfs, even with the overlapping telluric absorption.  
For this sight line, we use an analogous method to \cite{gilbert06} to weed out 
stars that are Milky Way dwarfs and therefore, not associated with And~X.  
The major difference in our case is that we are not targeting a kinematically 
diffuse M31 halo population, but rather a cold spike representing And~X.  
This difference is applied to the \cite{gilbert06} formalism by adjusting 
the ``training set'' to represent a Gaussian with $\sigma_v$ = 20~km~s$^{-1}$, 
centered on the rough velocity of And~X (we used $-$170~km~s$^{-1}$ for this).  The 
resulting classification of a member star vs non-member dwarf is not sensitively 
dependent on the choice of these parameters, as long as the initial guesses are 
not too restrictive. In this refined method, we have also adjusted the distance of 
the population to And X's known value (e.g., in computing photometric 
metallicities, see \S\,\ref{metallicity}).

With this initial classification, we find 28 stars that can be considered 
giants in this data set. However, a small number of these stars are expected 
to be M31 field halo members, unassociated with And~X.  To ensure as secure a 
sample as possible, even at the expense of possibly losing one or two member 
stars, we further eliminate four stars from this group of 28.  These stars 
all have velocities that are inconsistent at $>$3$\sigma$ with the mean of the 
remaining And~X sample, {\it and} three of them are also located more than 
three core-radii (and half the tidal radius) from the center of the dSph.  These 
stars, likely M31 field 
halo stars or possibly some residual Milky Way dwarf contaminants, have 
velocities of $v_{\rm rad}$ = $-$108.7~km~s$^{-1}$, $-$122.4~km~s$^{-1}$, 
$-$185.5~km~s$^{-1}$, and $-$209.5~km~s$^{-1}$.  As a final cut, we eliminate 
two more objects with radial velocities that fall within the range of And~X member 
stars, $v_{\rm rad}$ = $-$158.6~km~s$^{-1}$ and $v_{\rm rad}$ = $-$163.2~km~s$^{-1}$.  
The first object is also located more than three core-radii from the center of And~X, 
and is both brighter and bluer than the expected tip of the red giant branch 
on the CMD (see next section).  The second object is much redder than the red 
giant branch of And~X and is the only remaining star 
showing evidence for both Na\,{\sc i} and K\,{\sc i} absorption features.  
As we show below, And~X's radial velocity is almost exactly equal to the 
radial velocity of this star, and therefore our likelihood measurement was 
artificially biased into classifying this object as a member star.  We stress 
that the mean velocity and velocity dispersion measured for And~X are 
essentially unchanged if these two stars are left into the final sample, 
differing by $<$0.4~km~s$^{-1}$ and $<$0.2~km~s$^{-1}$ respectively (much smaller 
than the error bars, see below).

The final velocity histogram of the 22 member stars of And~X is shown as 
a solid line in Figure~\ref{fig:d10cmdvel2} (top).  The same plot 
over a finer velocity range and centered on the galaxy's mean velocity, 
where the small ticks are each 2.5~km~s$^{-1}$, is shown in the inset.  
Based on this secure sample of stars, the weighted mean radial velocity 
of And~X is $v_{\rm rad}$ = $-$163.8 $\pm$ 1.2~km~s$^{-1}$.  

\subsubsection{Color-Magnitude Diagram}\label{cmd}

The CMD for the stars along our And~X sight line is presented 
in Figure~\ref{fig:d10cmdvel2} (bottom).  This includes only 
those stars for which we measured a reliable velocity.  The open 
circles mark those objects that we determined to be non-members 
of And~X, most of which are Milky Way dwarfs.  As expected, these 
are found over a broad color range in the CMD.  The filled circles 
designate the confirmed And~X members and the open circles 
with crosses signify the six stars that were further rejected 
from membership as discussed above.

The filled circles in Figure~\ref{fig:d10cmdvel2} (bottom) are distributed 
in a narrow sequence of the CMD, signifying the red giant branch of And~X.  
To demonstrate this, we overlay three solid (dashed) curves on the CMD 
representing theoretical 
isochrones near the red giant tip with an age of 12~Gyr and [$\alpha$/Fe] 
= 0.0 ([$\alpha$/Fe] = $+$0.3) \citep{vandenberg06}.  The metallicities of 
these models are [Fe/H] = $\rm-$2.31 (left), [Fe/H] = $\rm-$1.31 (middle), 
and [Fe/H] = $\rm-$0.83 (right).  For this comparison, we have shifted the 
isochrones to a distance modulus of ($m - M$)$_{\rm 0} \sim$ 24.23 and 
reddened them by $E(V-I)$ = 0.21 \citep{zucker07}.  Clearly, the And~X 
stars are in good agreement with these metal-poor isochrones with the 
exception of the faintest stars with $I_{\rm 0}$ $\gtrsim$ 21.5.  Four of these 
faintest objects fall significantly redder than the dominant sequence, 
and one star falls bluer than the most metal-poor isochrone by 
$\sim$0.1~magnitudes.  

\begin{figure*}
\begin{center}
\leavevmode
\includegraphics[width=12cm, angle=270]{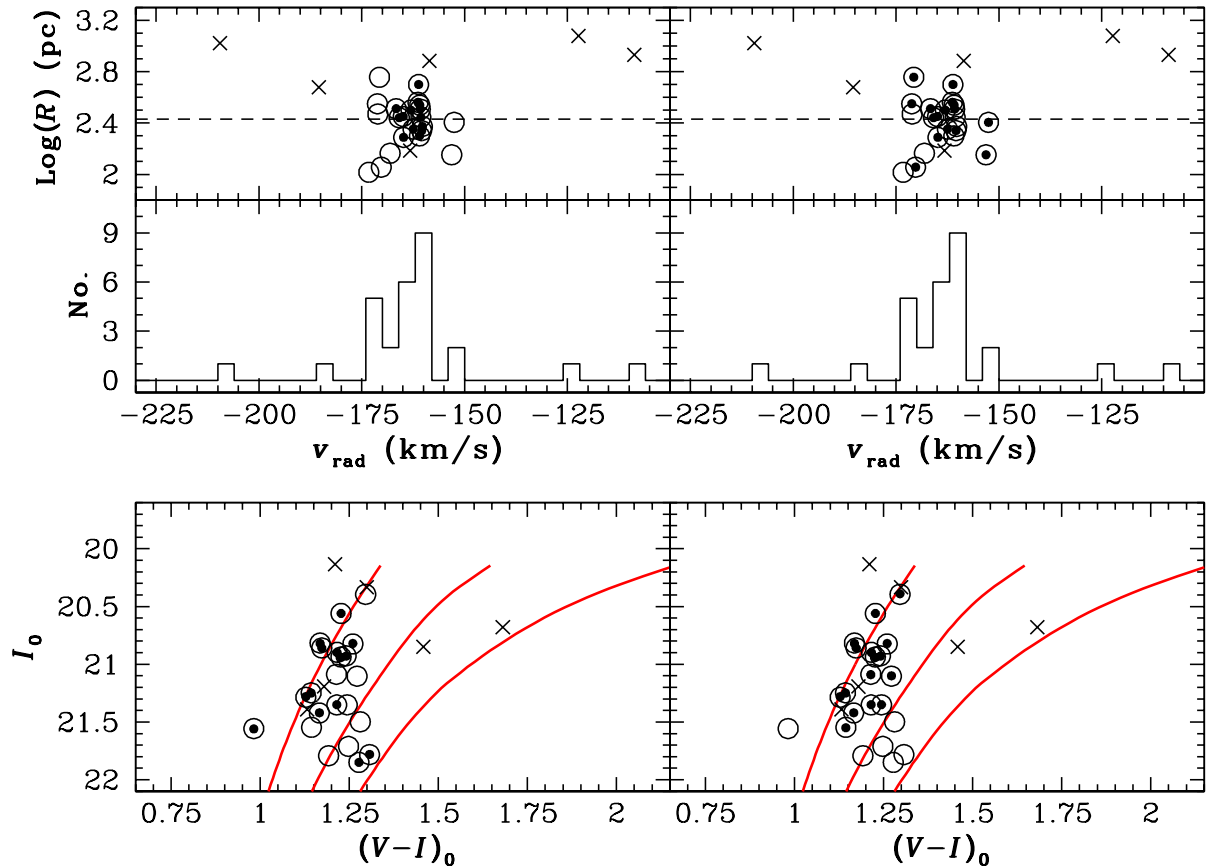}
\end{center}
\caption{A diagnostic plot comparing specific characteristics of stars along the 
line of sight, including their radial distance from the center of And~X vs the radial 
velocity (top panels, dashed line marks the core radius), a radial velocity histogram (middle panels), and the location 
of these stars on the CMD (bottom panels).  The open circles denote our 22 member stars of And~X, 
the open circles enclosing filled circles are the specific selections being analyzed (see below), and 
the crosses mark the six objects that are eliminated from the sample as discussed in \S\,\ref{velocity}.  
First, in the left panel, we isolate a core group of stars that have velocities in excellent agreement 
with the peak in the velocity histogram (formally within 3$\sigma$ of the measured And~X mean radial 
velocity of $v_{\rm rad}$ = $-$163.8 $\pm$ 1.2~km~s$^{-1}$).  (These dotted circles are difficult to 
discern on the top panel as they all reside in a narrow velocity range.)  The distribution of these 
stars on the CMD in the bottom panel illustrates that the stars with blue and red 
colors at the faint end of our sample are in fact not outliers in the velocity diagnostic.  
In the right hand panels, the stars are selected to reside within the tight red giant 
branch sequence (excluding the fainter stars) and the resulting distribution in the 
radius vs velocity plane does in fact include some of the members in the wings of the 
histogram.  These comparisons illustrate that the confirmed sample of stars is robust.}
\label{fig:diag}
\end{figure*}

The default photometry for 20 of our final member stars is derived from the 
William Herschel Telescope observations, whereas the remaining 
two stars are drawn from the Nordic Optical Telescope, both of which have 
$I_{\rm 0}$ $>$ 21.5.  In both catalogs, the CMD of all stars that were 
measured on the imaging frames indicates a clear increase in the spread of the 
red giant branch at $I_{\rm 0}$ $\sim$ 21.5, and so we estimate that the errors 
on the colors of these faintest stars are at least $\sigma_{(V-I)}$ $\sim$ 
0.2~magnitudes.  If we only consider the Nordic Optical Telescope data, we also 
find that two of the four faintest stars fall redder than an extension of the 
obvious red giant branch.  We consider more carefully the use of these stars in our 
metallicity calculations in \S\,\ref{metallicity}.  For completeness, we 
point out that these fainter stars are members of And~X as defined above 
and therefore the measured mean velocity (and intrinsic velocity dispersion) 
should minimally vary if they are removed from the sample.  To test this, 
we confirmed shifts in the inferred mean velocity of $<$0.7~km~s$^{-1}$ and 
in the calculated velocity dispersion (see below) of $<$0.2~km~s$^{-1}$ if 
the sample is restricted to just the tight sequence of points that are blueward 
of the [Fe/H] = $\rm-$1.31 isochrone (also excluding the very blue star at 
($V-I$)$_{\rm 0}$ = 0.98).  The characteristics of stars within several 
diagnostics including their radial distance from the center of And~X, 
line of sight velocities, and location in the CMD are compared to one another 
in Figure~\ref{fig:diag} to illustrate the robustness of the membership criteria.

There are a few non-And~X member stars in our overall sample 
with heliocentric velocities of $-$300 to $-$350~km~s$^{-1}$.  These objects 
are unlikely to be Milky Way dwarfs based on their kinematics and, in fact, also 
show no signs of Na\,{\sc i} or K\,{\sc i} lines in their spectra.  The objects 
are all relatively blue ($V-I$)$_{\rm 0}$ $\leq$ 1.5 and therefore represent 
strong candidate (metal-poor) M31 halo red giants in this field.  This is an 
important result in itself as it represents the first detection of the 
metal-poor, stellar halo of M31 in the north-east quadrant.

\subsection{Intrinsic Velocity Dispersion and $M/L$ Ratio}\label{velocitydispersion}

We calculate the velocity dispersion of And~X by first accounting 
for the random and systematic velocity measurement errors as discussed in 
\S\,\ref{accuracy}.  The observed dispersion is the sum of this error and 
the true, intrinsic velocity dispersion of the galaxy.  Using the 
maximum-likelihood method described by \cite{walker06}, we derive 
$\sigma_v = 3.9 \pm 1.2$~km~s$^{-1}$ for the 22 And~X member stars.  
The galaxy therefore ranks as the second coldest (kinematically) dSph 
in the Local Group.  The current record belongs to the Milky Way 
satellite Leo~IV, for which \cite{simon07} derive $\sigma_v = 3.3 
\pm 1.7$~km~s$^{-1}$ based on 18 member stars.  We note that the 
measured dispersion of And~X is relatively insensitive to an (unknown) 
red giant binary fraction of 20 -- 30\%.  As \cite{olszewski96} have shown, 
unresolved binaries alone can inflate the measured velocity dispersion by 
as much 1.5~km~s$^{-1}$.  Such an effect would reduce our measured velocity 
dispersion from 3.9 to 3.6~km~s$^{-1}$, a small change relative to the 
error bar in our measurement.

Such a low velocity dispersion for And~X is almost a factor of two 
smaller than the ``floor'' suggested by \cite{gilmore07} at 
$\sim$7~km~s$^{-1}$, based on previously studied dSphs.  In their 
study of newly discovered dSphs, \cite{simon07} found four Milky Way 
satellites with velocity dispersions significantly less than this 
floor, Leo~IV, Coma Berenices, Canes Venatici~II, and 
Hercules.  However, all four of these systems are considered very 
low-luminosity galaxies, with integrated brightnesses of just $-$5.9 
$< M_V <$ $-$3.7, and therefore are different from the existing 
sample of brighter galaxies.  An intrinsic velocity dispersion of 
$\sigma_v = 3.9 \pm 1.2$~km~s$^{-1}$ for And~X not only implies 
that such systems exist outside of the Milky Way's dSphs, 
but also presents the first evidence that such kinematically cold 
populations may exist in much brighter galaxies (And~X has an 
observed luminosity of $M_V$ = $-$8.1 $\pm$ 0.5).

We can estimate the mass of And~X using the method described by 
\cite{illingworth76}, $M$ = 167*$\beta$$r_c$$\sigma^2$, where 
$\beta$ = 8 for dSphs and $\sigma$ is the central velocity dispersion.  
The mean radius of our 22 member stars is equal to the core radius, 
and therefore the central velocity dispersion that is appropriate 
for this equation may be somewhat higher than the estimate from our 
sample.  Note that this method also explicitly assumes that 
the mass and light trace one another, which may not be correct 
(see below and \S\,\ref{discussion} for more discussion on this and 
other possibly incorrect assumptions).  The core radius of And~X is 271~pc 
\citep{zucker07}, and therefore we find $M$ = $5.4^{+3.8}_{-2.8} \times$ 
10$^6$~$M_\odot$.  The mass of the galaxy is reasonably similar to the group 
of four Milky Way dSphs with low intrinsic velocity dispersions discussed above, 
all of which have $M$ = 1 -- 7 $\times$ 10$^6$~$M_\odot$ calculated 
in the same way.  Of course, given that the integrated luminosity 
of And~X is much larger than these systems, the galaxy's mass-to-light 
ratio is smaller by a factor of 4 (in the case of Leo~IV) to 12 (in 
the case of Coma Berenices) than the Milky Way low-luminosity dSphs.  
Formally, we calculate the mass-to-light ratio of And~X to be 
$M/L_V$ = $37^{+26}_{-19}$ in solar units.  

We can also compare And~X's dark matter content to Milky Way 
satellites of comparable luminosity, calculated in a similar way.  
Both Draco and Ursa~Minor 
are metal-poor and have $M_V \sim$ $-$8.5, yet their mass-to-light 
ratios are $M/L_V \gtrsim$ 80 \citep{mateo98,odenkirchen01,lokas09}.  
Given the small difference in luminosity, these galaxies therefore contain 
roughly four times as much mass as And~X.  Two of the 
classical Milky Way dSphs with similar mass-to-light ratios to 
And~X are Carina and Sextans (both with $M/L_V$ = 30 -- 40), 
however these systems are $\sim$3$\times$ brighter than And~X.  
We discuss the possible implications of these results further in 
\S\,\ref{discussion}.

It is important to note the limitations of the method described 
above to calculate mass-to-light ratios, not just for And~X but 
also for many of the other Local Group dSphs.  A central velocity 
dispersion from a few dozen measurements alone offers degenerate 
constraints on the dark matter mass of a galaxy (e.g., Zentner \& 
Bullock 2003).  We have assumed that the galaxy is spherical, in 
dynamical equilibrium, and has an isotropic velocity dispersion, 
all of which may not be true \citep{simon07}.  As mentioned above, 
we have also assumed that mass traces light.

A more accurate measure of the amount, and nature, of dark matter in 
these systems requires a full velocity dispersion profile (e.g., 
Strigari et~al.\ 2007; 2008).  This modeling includes a maximum 
likelihood analysis and full marginalization over the dark matter 
halo profile shape and the stellar velocity anisotropy, and 
benefits from large samples of accurate radial velocity measurements 
(specifically out to large radii).  Of course, a different set of 
assumptions related to the Jeans mass modeling affects these 
measurements.  We discuss the prospects for achieving 
such data sets for the M31 system of dwarf satellites in \S\,\ref{discussion}.
For our And~X data set, detailed modeling along these lines will be presented 
in J.\ Wolf~(2009, in preparation).  Preliminary results suggest that the 
{\it total} mass (poorly constrained) out to the tidal radius in this galaxy 
is larger than the rough estimate given above.  However, the mass at the half 
light radius, with much better constraints, is in fact systematically lower 
than the Milky Way counterparts of similar luminosity as reported above 
(modeled the same way).  


\section{Chemical Abundances}\label{metallicity}

\subsection{Photometric and Spectroscopic Metallicities}

The present day chemical abundances of dSphs are important tracers of their 
integrated star formation histories.  The mean abundances, and 
abundance spreads, vary tremendously from one dwarf to another, which 
suggests that some galaxies were able to retain their enrichment 
products more efficiently than others.  The CMD presented in 
Figure~\ref{fig:d10cmdvel2} shows that And~X is a metal-poor 
system (the middle isochrone for 12~Gyr has [Fe/H] = $-$1.31).  We can measure the mean 
metallicity of the system by interpolating the positions of the stars on the CMD 
within a fine grid of stellar isochrones.  Over the range $-$2.31 $<$ [Fe/H] $<$ 
$-$0.83, the \cite{vandenberg06} models consist of a dozen such models.  For the 
few stars just to the left of the most metal-poor isochrone, we obtain an [Fe/H] 
value by slightly extrapolating the grid.  The resulting metallicity distribution 
function is shown in Figure~\ref{fig:d10met}, assuming purely old ages (see below) and 
both [$\alpha$/Fe] = 0.0 isochrones (solid) and alpha-enhanced models with 
[$\alpha$/Fe] = $+$0.3 (dashed).  The median metallicity of the 22 confirmed 
members of And~X is [Fe/H]$_{\rm phot}$ = $-$1.93 $\pm$ 0.11 for [$\alpha$/Fe] 
= 0.0 and [Fe/H]$_{\rm phot}$ = $-$2.13 $\pm$ 0.11 for [$\alpha$/Fe] = $+$0.3.  
The {\it mean} metallicities are consistent within the 1$\sigma$ uncertainties of 
these median values.  

The absolute photometric uncertainty in color for the bulk of our stars with $I_{\rm 0}$ $\lesssim$ 
21.5 is $\sim$0.05 mags, which translates to a metallicity error of $\sigma$[Fe/H] 
$\sim$ 0.20 at the bright, metal-poor end of the CMD.  The CMD of And~X members presented 
earlier in \S\,\ref{cmd} also illustrates that the uncertainty may be larger 
than this for the faintest stars in our member sample, with four of them falling significantly 
redder than the remaining sequence and one star lying much bluer than the most 
metal-poor isochrone.  If we eliminate these five stars from our metallicity 
analysis, the median and mean metallicity of And~X remains essentially the same, 
changing by $<$0.02 dex.  



\begin{figure}
\begin{center}
\leavevmode
\includegraphics[width=8.5cm]{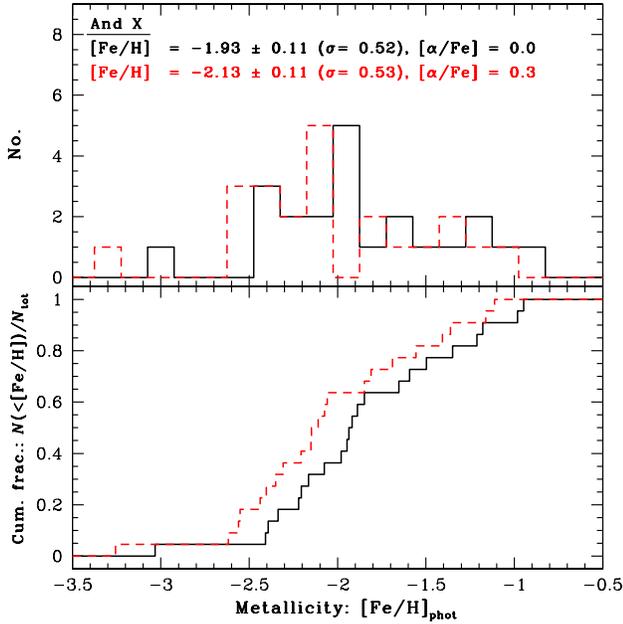}
\end{center}
\caption{The metallicity distribution function is calculated by interpolating 
the photometry of confirmed And~X stars within a grid of isochrones, assuming 
both [$\alpha$/Fe] = 0.0 (solid) and [$\alpha$/Fe] = $+$0.3 (dashed).  
The differential histogram is shown at the top and the cumulative distribution 
is shown in the bottom panel.  And~X is clearly a very metal-poor galaxy, 
with [Fe/H] = $-$1.9 to $-$2.1, depending on the assumed $\alpha$-abundance.  The 
intrinsic spread in chemical abundances is also large, $\sigma$ = 0.48 dex.  We compare 
these measurements to independently determined spectroscopic metallicities in 
Figure~\ref{fig:d10coadd}.}
\label{fig:d10met}
\end{figure}



\begin{figure}
\begin{center}
\leavevmode
\includegraphics[width=8.5cm]{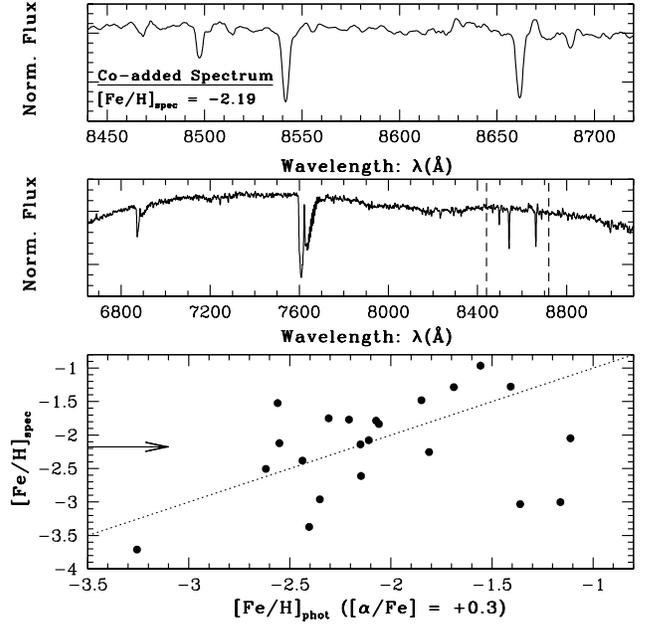}
\end{center}
\caption{{\it Bottom --} The photometric metallicities calculated in 
\S\,\ref{metallicity} (for the [$\alpha$/Fe] = $+$0.3 isochrones) are compared 
to independently measured spectroscopic estimates based on a calibration of 
the equivalent width of the Ca\,{\sc ii} triplet absorption lines.  The straight 
dotted line represents the 1:1 relation and the arrow marks the mean spectroscopic 
abundance, [Fe/H]$_{\rm spec}$ = $-$2.18 $\pm$ 0.15.  The middle panel exhibits 
the co-added spectrum of the 22 individual member red giants in And~X.  A zoomed 
view of this spectrum, over the region marked by the two vertical dashed lines is 
shown in the top panel.  This high signal-to-noise spectrum yields a 
metallicity of [Fe/H]$_{\rm spec}$ = $-$2.19, consistent with both the mean of 
the individual stars and also the photometric metallicity (see discussion in 
\S\,\ref{metallicity}).}
\label{fig:d10coadd}
\end{figure}


Our inferred metallicity of And~X is based on photometric measurements of 
confirmed member red giant branch stars.  This method assumes an input age 
and [$\alpha$/Fe].  Unfortunately, there is currently no direct constraint 
on either of these parameters for And~X.  For the age, such a measurement 
would require deeper imaging to detect the morphology of the horizontal branch 
or main-sequence turnoff.  We make the assumption that {\it most} of the 
stars in the galaxy are old based on both the fact that this is true for 
most other dSphs whose star formation histories have been measured (e.g., 
Mateo 1998; Harbeck et~al.\ 2001), and because our deeper CMD does not show any 
evidence for a pronounced population of thermally pulsating asymptotic giant branch 
stars or a vertical red clump (see Nordic Optical Telescope data in Figure~3 of Zucker 
et al.\ 2007).  Even if this assumption is wrong, our metallicity scale will not be 
drastically affected.  For a shift in age from 12~Gyr to 6~Gyr, the mean 
metallicity of a (metal-poor) star near the tip of the red giant branch 
would shift by only $+$0.25 dex.  Similar to the uncertainty in age, the 
detailed chemical abundances of 
these stars are not known.  For the Milky Way dSphs, [$\alpha$/Fe] can 
be measured by analyzing high resolution spectra of individual red giants.  
Studies of these stars in several Galactic dSphs (e.g., Draco, Sextans, 
and Ursa Minor) initially suggested that these objects are slightly enhanced in 
$\alpha$-elements, 0.02 $\lesssim$ [$\alpha$/Fe] $\lesssim$ 0.13 
\citep{shetrone01}, however recent work has found solar and/or slightly subsolar 
values (e.g., Cohen \& Huang 2009 and references therein).  As Figure~\ref{fig:d10met} 
demonstrates, such enhancement leads to a mean abundance that is {\it more metal-poor} than 
measured using our standard set of solar-scaled isochrones.  Taken together, 
although it is clear that we currently have no {\it direct} constraints on 
either age or $\alpha$-enhancement in And~X, we can qualitatively say that 
any reasonable offsets of these parameters from the values assumed would not 
cause large changes in our metallicity estimates.

An alternative method to determine the metallicity of red giant 
stars is to analyze their spectral features.  Unfortunately, at the distance 
of And~X the signal-to-noise of the individual spectra are too low to 
accurately characterize many absorption lines.  The only spectral features 
that can be used to estimate [Fe/H] from these data are the Ca\,{\sc ii} 
triplet at 8498, 8542, and $\rm8662\AA$ (see Rutledge, Hesser, \& Stetson 1997; 
Rutledge et~al.\ 1997 for more information).  Still, these lines reside in a 
region of the spectrum that contains many 
night sky lines and therefore the measured equivalent widths have large errors and 
the scatter in the spectroscopic metallicity measurements will be inflated 
(see e.g., Kalirai et~al.\ 2006a).  Although it is difficult to characterize the 
uncertainties, we nevertheless show the comparison of our measured [Fe/H]$_{\rm phot}$ 
values to [Fe/H]$_{\rm spec}$ measurements calculated in this way in 
Figure~\ref{fig:d10coadd} ({\it bottom}).  The empirical relation used in this calculation 
is derived using Galactic globular clusters and therefore the comparison has been 
made to the photometric metallicities assuming the [$\alpha$/Fe] = $+$0.3 
isochrones.  We also point out that most globular clusters are more metal-rich 
than [Fe/H] $\sim$ $-$2.2, and therefore this calibration is extrapolated for 
very metal-poor stars. \cite{koch08b} demonstrate that this leads to an underestimate 
of the true metallicity by several tenths of dex for very metal-poor stars.  
Further details on this empirical calibration can be found in Rutledge 
et~al.\footnote{We have assumed the luminosity of And~X's horizontal branch 
to be $V_{\rm HB}$ = 25.0.}.  The dotted line represents the 1:1 relation, 
and in general, the two independently determined metallicity values are found 
to agree well, although the scatter is large.  The arrow marks the mean metallicity 
of the spectroscopic measurements, [Fe/H]$_{\rm spec}$ = $-$2.18 $\pm$ 0.15, 
which is very similar to our photometric metallicity for [$\alpha$/Fe] = $+$0.3 
([Fe/H]$_{\rm phot}$ = $-$2.13 $\pm$ 0.11).

Figure~\ref{fig:d10coadd} ({\it bottom}) can also be used to take a closer look 
at the five stars with possible poor photometry discussed above.  Interestingly, the one 
very blue star in our CMD with [Fe/H]$_{\rm phot}$ = $-$3.26 for [$\alpha$/Fe] 
= $+$0.3 is also found to be spectroscopically metal-poor.  This star therefore 
represents a very interesting, and rare, metal-poor candidate in a dSph.  
\cite{helmi06} compare the metallicity distribution function of Galactic dSphs 
with the Galactic halo, and find a lack of [Fe/H] $<$ $-$3 stars in the dSph 
population, suggesting that these satellites are different from the building 
blocks of the halo (but, also see Kirby et~al.\ 2008).  A higher signal-to-noise 
spectrum of this star can yield a more accurate spectroscopic metallicity.   Contrary 
to what we find for this bluer star, the spectroscopic metallicities do not agree 
with the photometric estimates for three of the four faint stars that appear to be redder 
than the brighter, well defined red giant branch in Figure~\ref{fig:d10cmdvel2}.  
These three stars are shown at the metal-rich end of the photometric scale on 
Figure~\ref{fig:d10coadd}, and clearly fall well below the 1:1 relation.  This 
comparison provides further evidence that the colors of these four stars are 
likely incorrect by at least $\sim$0.2 magnitudes, and that the actual metallicities 
of the stars are in fact more metal-poor than we have determined from the isochrone 
analysis.  Of course, if we just eliminate these three stars from the analysis, keeping 
the one blue star in our sample, the mean metallicity of And~X becomes more metal-poor 
by $\sim$0.1 dex compared to the values given above.

In the top two panels of Figure~\ref{fig:d10coadd} we present a high signal-to-noise co-added 
spectrum of the 22 individual spectra for confirmed And~X members.  This was constructed 
by weighting the individual spectra by the inverse variance squared to minimize the 
contribution from features affected by bad pixels or residual night sky lines.  
The middle panel shows the full spectrum and the top panel shows the spectrum in the 
wavelength range of the Ca\,{\sc ii} triplet (i.e., the dashed lines in the middle panel).  
The Ca\,{\sc ii} triplet lines in this co-added spectrum are well defined and 
therefore the equivalent widths can be measured to greater precision than in the 
individual spectra.  Using the same method as above on this single co-added spectrum, we 
find a value for the spectroscopic metallicity of [Fe/H]$_{\rm spec}$ = $-$2.19, 
essentially identical to the mean of the individual measurements.  Given the method 
used to co-add the individual spectra, we do not have any information on the spread 
in abundances specifically from this measurement.

To summarize, we have used two independent diagnostics to determine the chemical abundance 
of And~X.  Each of these two methods has different disadvantages.  For example, the 
photometric measurement is sensitive to the quality of the photometry and the 
spectroscopic metallicity is affected by the low signal-to-noise of our spectra.  
Nevertheless, the results from the two methods are found to be in good agreement.  As 
we do not know whether stars in this galaxy are $\alpha$-enhanced, we adopt the photometric 
metallicity for [$\alpha$/Fe] = 0.0 as our best value, [Fe/H]$_{\rm phot}$ = $-$1.93 
$\pm$ 0.11, and note that the metallicity would be more metal-poor by $\sim$0.1 dex if 
the galaxy is significantly enhanced in $\alpha$-elements (i.e., [$\alpha$/Fe] = $+$0.3).  

Based on these calculations, And~X is among the most metal-poor nearby dSph galaxies 
with $M_V$ $\lesssim$ $-$8 (e.g., Table~1 in Grebel, Gallagher, \& Harbeck 2003).  
In terms of the well-known correlation between the luminosity of a dwarf galaxy and 
its metallicity over this range \citep{aaronson86,mateo98}, And~X sits near the previously 
established relation for dSphs, anchoring the faint, metal-poor end.  A strong deviation 
from this relation (e.g., a dwarf that is too faint for its metallicity) may suggest that the 
galaxy lost some of its stellar component through tidal disruption.  Our metallicity 
analysis does not provide any strong evidence that And~X's present day population has 
been affected in this way.  Of course, \cite{simon07} have shown that the metallicities 
of the newly discovered SDSS low-luminosity galaxies with $-$8 $\lesssim$ $M_V$ 
$\lesssim$ $-$3.5 may be even more metal-poor than [Fe/H] $\sim$ $-$2 (see also Kirby 
et~al.\ 2008).

Derived parameters for each of the 22 confirmed red giants in And~X are presented in Table~2.

\subsection{Metallicity Spread and Radial Metallicity Gradient}


\begin{figure}
\begin{center}
\leavevmode
\includegraphics[width=8.5cm]{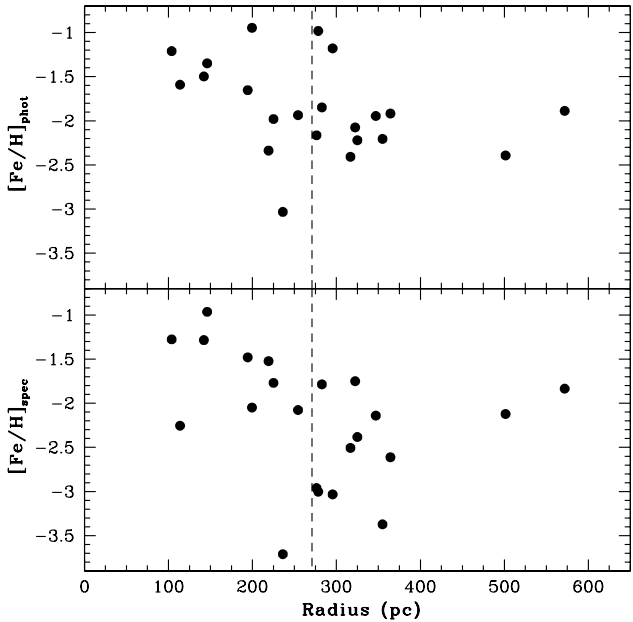}
\end{center}
\caption{The photometric ({\it top}) and spectroscopic ({\it bottom}) 
metallicities indicate the presence of a radial abundance gradient in And~X.  
The inner regions of the dSph are clearly more metal-rich than stars near the 
core radius ($r_{\rm c}$ = 271~pc, dashed line).}
\label{fig:d10radialtrend4}
\end{figure}


The total measured abundance dispersion for And~X is relatively large, $\sigma$([Fe/H]$_{\rm phot}$) 
= 0.52.  This dispersion represents a combination of the true intrinsic dispersion of the 
galaxy and the scatter introduced by photometric errors.  If we subtract off the 
error term in quadrature ($\sim$0.2 dex), the resulting intrinsic abundance dispersion is 
calculated to be $\sigma$([Fe/H]$_{\rm phot}$) = 0.48.  This is not unlike typical values found 
for Milky Way dSphs \citep{mateo98} and suggests that And~X was able to retain much of its 
enrichment products.  A full interpretation of this will of course benefit greatly from 
knowledge of the star formation history of the galaxy, as measured from deeper photometric 
observations.  For example, if the galaxy shows evidence for an extended epoch of star 
formation as seen in Fornax and Leo~I, then a small fraction of the spread in color on the 
upper red giant branch can be explained by age differences.  We also note that with a 
sample of just 22 stars, our results are mildly sensitive to any extreme outliers in the 
metallicity distribution.  If we remove the three objects with discrepant photometric vs 
spectroscopic metallicities discussed above, the dispersion decreases to 
$\sigma$([Fe/H]$_{\rm phot}$) = 0.38.

In Figure~\ref{fig:d10radialtrend4} we illustrate the photometric ({\it top panel}) 
and spectroscopic ({\it bottom panel}) metallicities of stars as a function of 
their distance from the center of And~X.  The core radius of the galaxy, $r_{\rm c}$ = 271~pc 
is marked with a dashed line.  There is clear evidence of a radial metallicity gradient in 
And X, with more metal-rich stars being centrally located.  These innermost stars have 
[Fe/H]$_{\rm phot} \sim$ $-$1.4 which gradually decreases to [Fe/H]$_{\rm phot} \lesssim$ $-$2.0 
near, and beyond, the core radius.  We also note that two of the faintest four stars in our data set 
with redder $V-I$ colors are located at $r$ $\sim$ 300~pc, and therefore these stars do not drive 
the observed gradient.  

In order to distinguish between different formation scenarios of this galaxy, the abundance 
gradient should be verified using deeper photometry to detect a morphological change in the 
horizontal branch of the galaxy as a function of radius.  The 
result could also suggest that an age gradient is the driver of the observed metallicity 
gradient, which can be tested with photometric studies of the main-sequence turnoff of 
And~X. \cite{harbeck01} note that of the other M31 dSphs that have been studied photometrically 
down to the horizontal branch, only And~I and VI show an obvious population gradient.


\begin{table*}
\begin{center}
\caption{}
\begin{tabular}{lccccccccc}
\hline
\hline
\multicolumn{1}{c}{Mask} & \multicolumn{1}{c}{$\alpha_{J2000}$} & 
\multicolumn{1}{c}{$\delta_{J2000}$} & \multicolumn{1}{c}{$I_{\rm 0}$} & 
\multicolumn{1}{c}{($V-I$)$_{\rm 0}$} & \multicolumn{1}{c}{$r$ (pc)} & 
\multicolumn{1}{c}{$v_{\rm rad}$ (km~s$^{-1}$)} & \multicolumn{1}{c}{[Fe/H]$_{\rm phot}$} & 
\multicolumn{1}{c}{[Fe/H]$_{\rm phot}$} & \multicolumn{1}{c}{[Fe/H]$_{\rm spec}$} \\ 
& & & & & & & 
\multicolumn{1}{c}{([$\alpha$/Fe] = 0.0)} & \multicolumn{1}{c}{([$\alpha$/Fe] = 0.3)}
\\ 

\hline
d10$\_$1 & 01:06:30.83 &  $+$44:47:50.2  &  20.39  &  1.30  &  355  &  $-$171.2 $\pm$ 3.1  &  $-$2.20  & $-$2.40  &  $-$3.37 \\
d10$\_$1 & 01:06:28.43 &  $+$44:47:11.9  &  20.56  &  1.23  &  501  &  $-$161.2 $\pm$ 3.2  &  $-$2.39  & $-$2.55  &  $-$2.12 \\
d10$\_$1 & 01:06:29.79 &  $+$44:48:25.1  &  20.82  &  1.26  &  347  &  $-$160.9 $\pm$ 4.4  &  $-$1.95  & $-$2.15  &  $-$2.14 \\
d10$\_$2 & 01:06:46.08 &  $+$44:47:47.0  &  20.82  &  1.17  &  317  &  $-$163.0 $\pm$ 4.2  &  $-$2.41  & $-$2.62  &  $-$2.51 \\
d10$\_$2 & 01:06:38.74 &  $+$44:47:40.3  &  20.86  &  1.17  &  219  &  $-$160.4 $\pm$ 2.7  &  $-$2.34  & $-$2.56  &  $-$1.52 \\
d10$\_$2 & 01:06:34.60 &  $+$44:47:23.4  &  20.90  &  1.22  &  322  &  $-$160.8 $\pm$ 4.2  &  $-$2.08  & $-$2.31  &  $-$1.75 \\
d10$\_$1 & 01:06:41.01 &  $+$44:50:30.0  &  20.93  &  1.24  &  364  &  $-$161.3 $\pm$ 5.8  &  $-$1.92  & $-$2.15  &  $-$2.61 \\
d10$\_$2 & 01:06:32.98 &  $+$44:48:43.6  &  20.94  &  1.23  &  225  &  $-$162.5 $\pm$ 3.4  &  $-$1.98  & $-$2.21  &  $-$1.77 \\
d10$\_$2 & 01:06:34.21 &  $+$44:47:51.9  &  21.09  &  1.21  &  254  &  $-$152.5 $\pm$ 3.1  &  $-$1.94  & $-$2.11  &  $-$2.08 \\
d10$\_$2 & 01:06:36.48 &  $+$44:48:27.8  &  21.10  &  1.27  &  114  &  $-$170.3 $\pm$ 3.1  &  $-$1.59  & $-$1.81  &  $-$2.25 \\
d10$\_$2 & 01:06:45.98 &  $+$44:48:07.2  &  21.25  &  1.14  &  276  &  $-$165.9 $\pm$ 4.3  &  $-$2.16  & $-$2.35  &  $-$2.96 \\
d10$\_$2 & 01:06:34.27 &  $+$44:47:24.7  &  21.28  &  1.13  &  325  &  $-$166.6 $\pm$ 3.5  &  $-$2.22  & $-$2.44  &  $-$2.38 \\
d10$\_$1 & 01:06:35.21 &  $+$44:48:06.3  &  21.35  &  1.22  &  194  &  $-$164.8 $\pm$ 2.7  &  $-$1.65  & $-$1.85  &  $-$1.48 \\
d10$\_$2 & 01:06:35.70 &  $+$44:48:25.6  &  21.35  &  1.24  &  142  &  $-$153.1 $\pm$ 4.9  &  $-$1.50  & $-$1.69  &  $-$1.28 \\
d10$\_$2 & 01:06:32.34 &  $+$44:49:24.4  &  21.42  &  1.17  &  283  &  $-$164.9 $\pm$ 3.4  &  $-$1.85  & $-$2.07  &  $-$1.79 \\
d10$\_$2 & 01:06:36.37 &  $+$44:48:39.5  &  21.50  &  1.28  &  104  &  $-$173.3 $\pm$ 4.3  &  $-$1.21  & $-$1.41  &  $-$1.28 \\
d10$\_$1 & 01:06:24.40 &  $+$44:47:46.0  &  21.55  &  1.14  &  572  &  $-$170.7 $\pm$ 6.7  &  $-$1.89  & $-$2.06  &  $-$1.84 \\
d10$\_$1 & 01:06:32.70 &  $+$44:48:51.0  &  21.56  &  0.98  &  236  &  $-$160.3 $\pm$ 10.1 &  $-$3.03  & $-$3.26  &  $-$3.71 \\
d10$\_$2 & 01:06:44.32 &  $+$44:47:36.9  &  21.71  &  1.29  &  296  &  $-$171.1 $\pm$ 5.1  &  $-$1.18  & $-$1.36  &  $-$3.03 \\
d10$\_$2 & 01:06:33.85 &  $+$44:48:30.4  &  21.78  &  1.31  &  200  &  $-$161.0 $\pm$ 9.4  &  $-$0.95  & $-$1.11  &  $-$2.05 \\
d10$\_$1 & 01:06:35.49 &  $+$44:48:27.6  &  21.79  &  1.19  &  146  &  $-$168.1 $\pm$ 12.2 &  $-$1.35  & $-$1.56  &  $-$0.96 \\
d10$\_$1 & 01:06:32.69 &  $+$44:48:01.0  &  21.85  &  1.28  &  278  &  $-$160.6 $\pm$ 5.8  &  $-$0.98  & $-$1.16  &  $-$3.00 \\
\hline
\end{tabular}
\label{table2}
\end{center}
\end{table*}





\section{Discussion and Summary} \label{discussion}

And~X has an integrated magnitude ($M_V$ = $-$8.1 $\pm$ 0.5) that bridges 
the luminosity of the Local Group dwarf galaxies known prior to SDSS 
($-$15.5 $< M_V < -$8.5), and those since discovered ($-$8 $< M_V < -$3.5).  
Over this factor of nearly 10$^{5}$ in luminosity (more than 12 magnitudes), these 
galaxies show a remarkable trend with more more luminous dwarfs also being 
more metal-rich (see, e.g., Figure~11 in Simon \& Geha 2007).  We have shown -- 
using both photometric and spectroscopic diagnostics -- that the stellar populations 
in And~X are metal-poor, with [Fe/H] = $-$1.9 to $-$2.1, depending on the 
method adopted and/or assumption of [$\alpha$/Fe].  The galaxy therefore 
anchors the low luminosity end of the classical satellites, and has a mean 
metallicity consistent with the established trend.  When we also 
include the new SDSS galaxies, And~X is well placed within the overall 
sequence of data points and does not appear to be underluminous relative 
to its metallicity (which could have suggested evidence of 
tidal stripping).

Our kinematic analysis of And~X shows that the galaxy is moving with a 
heliocentric radial velocity of $v_{\rm rad} = -163.8 \pm 1.2$~km~s$^{-1}$, 
and has a very low intrinsic velocity dispersion, $\sigma_v = 3.9 \pm 
1.2$~km~s$^{-1}$.  Compared to similar luminosity Milky Way dSphs such as 
Carina, Draco, Ursa Minor, and Canes Venatici I, And~X's intrinsic velocity 
dispersion is a factor of 1.7 -- 2.4 smaller.  Although its size is larger 
than Carina, Draco, and Ursa Minor by a similar factor (see below), the ratio of 
$r_c$$\sigma^2$ comes out lower.  We use this to estimate the mass of And~X 
(assuming mass follows light) and find $M$ = $5.4^{+3.8}_{-2.8} \times$ 
10$^6$~$M_\odot$, implying a mass-to-light ratio of $M/L_V$ = 
$37^{+26}_{-19}$.  The dark matter mass of And~X is therefore approximately 
a factor of four lower than Milky Way dSphs with comparable, but slightly 
higher, luminosity (see \S\,\ref{velocitydispersion}).  

It is difficult to draw strong conclusions on the 
possible implications of the mass difference between And~X and similar 
Milky Way satellites.  The differences may simply reflect variations of in 
situ processes, in which case And~X may be an oddball with respect to 
the small comparison sample of galaxies at similar luminosities.  Alternatively, 
if And~X is a typical representative of the M31 dSph sample, the observed differences 
may have been possibly shaped by intrinsic differences in the overall properties of 
the Milky Way and M31 halos.  For example, the past interaction history of 
these dSphs with their hosts are likely different in the two systems as M31 appears to 
have a more violent accretion history \citep{hammer07}.  Whether or not this could 
lead to preferentially more stripping of the dark matter halos of the surviving 
satellites such as And~X is not certain, especially within the visible radius of the 
galaxy (e.g., stars should also be stripped then).  Hints of overall global differences 
between M31 and the Milky Way have already been characterized, on several fronts.  For 
example, M31 contains many more globular clusters than the Milky Way (e.g., Barmby \& 
Huchra 2001), M31's stellar density at 20~kpc along the minor axis is 10$\times$ 
higher than the Milky Way \citep{reitzel98}, and M31's halo contains a vast amount 
of substructure extending out to very large radii \citep{ibata07}.  Further 
differences are seen in the overall age, metallicity, and surface brightness 
distributions, as characterized in our overall SPLASH Survey.  For example, M31's 
halo contains an appreciable fraction of stars with ages less than 10~Gyr out to 
20~kpc along the minor axis \citep{brown07}, and the radius at which the metal-poor 
halo dominates the metal-rich inner spheroid is much further out in M31 than in 
the Milky Way \citep{guh05,kalirai06a}.

Interestingly, there is another piece of evidence suggesting systematic differences 
between the dSph populations of the Milky Way and M31. As we mentioned in 
\S\,\ref{intro}, \cite{mcconnachie06} analyzed photometric data of a half dozen 
M31 dSphs and found that at a fixed luminosity, these satellites are larger by a 
factor of two and have fainter central surface brightnesses than Milky Way 
satellites. And~X is not included in that sample, however it does appear to follow 
the same trend.  The half light radius is 330~pc and tidal radius is 1500~pc, 
both much larger than the corresponding sizes of similar luminosity 
Milky Way satellites.  Yet, as discussed above, the very low intrinsic 
velocity dispersion of And~X more than compensates for its larger size in 
yielding a total mass that is {\it lower} than these Milky Way satellites.  

And~X's mass measurement depends critically on our assumptions related to how the 
dark matter mass links to the stellar distribution; namely, we have adopted the formalism 
that the mass traces light.  A different interpretation of the kinematics 
would result from an overall different spatial segregation.  For example, Penarrubia, 
McConnachie, \& Navarro (2007) compute models in which the kinematics are dictated 
by the ratio of the King core radius (stellar) to the NFW scale radius (dark matter), 
and find that a more concentrated distribution exhibits a lower velocity dispersion.  
With this assumption, less dark matter mass is needed in the larger galaxies to 
explain the same observed velocity dispersion.  Penarrubia et~al.\ go on to 
predict that, assuming M31 and Milky Way dSphs have the same overall dark matter 
mass, the velocity dispersions of the M31 satellites should be larger than the 
Milky Way satellites.  Based on our sample of a single M31 dSph, our results do not 
agree with this prediction.

There are currently 17 known M31 dSphs, And~I -- III, V -- VII, IX -- XVII, and 
XIX -- XX, the last ten of which have been discovered in the past three years 
\citep{zucker04b,zucker07,martin06,majewski07,ibata07,irwin08,mcconnachie08}.  
(The three systems not included, And~IV, VIII and XVIII, are background or 
ambiguous objects originally identified as M31 dSphs.)  Radial velocities of at 
least a few individual red giants have been measured in the first six of these 
galaxies \citep{cote99,guh00,evans00}.  Unfortunately, with the exception of 
stars in And~II and And~VII, which were 
targeted with HIRES on Keck, the data were collected at both low resolution and 
low signal-to-noise, resulting in large velocity uncertainties, and therefore 
the intrinsic velocity dispersion of the satellites were not resolved.  
For the two galaxies observed with Keck/HIRES, the central velocity 
dispersion was estimated to be $\sim$9~km~s$^{-1}$ in both galaxies, based 
on seven stars in And~II \citep{cote99} and 18 stars in And~VII 
\citep{guh00,kalirai09}.  For the newly discovered SDSS satellite And~IX 
\citep{zucker04b}, \cite{chapman05} present a Keck/DEIMOS study 
establishing just five member stars within the inner 1.4$'$ of the galaxy 
and possibly another six or seven members at larger radius.  As the authors 
note, their results are very sensitive to the adopted membership 
criteria.  The inclusion of just one star in the outer bin inflates the 
velocity dispersion by a factor of almost two, and therefore we can not 
reliably use this measurement until further data are obtained.  \cite{majewski07} 
also present radial velocities for stars in the newly discovered galaxy And~XIV, 
which we will analyze in Kalirai et~al.\ (2009).

Clearly, all of these M31 dSphs need to be re-targeted to establish accurate 
kinematics based on larger numbers of stars.  In addition, the newly detected 
objects that have not yet been looked at need to be observed.  Our team has 
now begun a systematic survey to compile this data set, the first results 
from which will include an analysis of 50 -- 100 accurate radial velocities in 
each of And~I, II, and III.  We will also combine the results from the study of 
these three galaxies with a more detailed analysis of the results presented in 
\cite{guh00} for And~VII, \cite{majewski07} for And~XIV, and of course also 
include the present work on And~X.  As discussed earlier, future modeling of 
these dSphs will also incorporate advanced techniques such as those presented 
by \cite{strigari08} for Milky Way galaxies, allowing mass profiles to be 
constructed for each satellite.  In this way, the dark matter masses of dSph 
members of the Milky 
Way and M31 can be compared out to the same physical radii.  With a sample of 
six galaxies, these comparisons can also be made over a large range in 
luminosity and galactocentric radius, thereby providing more leverage on 
understanding the sources of any observed differences.

A {\it detailed} understanding of the dark matter mass profiles of the M31 
dSphs will require significantly larger data sets involving accurate radial 
velocities of several hundred stars in each galaxy.  Such data can provide 
information on how luminosity maps to dark matter subhalo mass and also shed 
light on the validity of alternative dark matter candidates such as ``warm 
dark matter''.  These types of detailed studies are just now underway for 
several of the Milky Way dSphs (e.g., Strigari et~al.\ 2006) and will soon 
be undertaken in the nearest, most luminous dSphs of M31.  To this effort, 
we have now obtained the first wave of the data required to make these 
measurements; an analysis of $\sim$500 individual radial velocities in 
the luminous dSph And~II is underway.


\acknowledgments

We gratefully acknowledge E.~Kirby for help with calculating photometric and 
spectroscopic metallicities, and M.~Schirmer and T.~Pursimo for assistance in 
reducing the William Herschel and Nordic Optical Telescope data.  We wish to 
also thank J.~Wolf for very helpful extensive discussions on the mass modeling 
of these kinematic data.  Finally, we wish to thank an anonymous referee for 
many suggestions that have improved this paper.

JSK's research is supported in part by a grant from the STScI Director's 
Discretionary Research Fund, and was supported by NASA through Hubble Fellowship 
grant HF-01185.01-A, awarded by the Space Telescope Science Institute, which is operated by the 
Association of Universities for Research in Astronomy, Incorporated, under 
NASA contract NAS5-26555.  This project was also supported by NASA/STScI 
grant GO-10265.02 (JSK and PG).  DBZ acknowledges support from a National 
Science Foundation International Postdoctoral Fellowship.  

Funding for the SDSS and SDSS-II has been provided by the Alfred P.\ Sloan 
Foundation, the Participating Institutions, the National Science Foundation, the 
U.S.\ Department of Energy, the National Aeronautics and Space Administration, 
the Japanese Monbukagakusho, the Max Planck Society, and the Higher Education 
Funding Council for England. The SDSS Web Site is \texttt{http://www.sdss.org/}.

The SDSS is managed by the Astrophysical Research Consortium for the Participating 
Institutions.  The Participating Institutions are the American Museum of 
Natural History, Astrophysical Institute Potsdam, University of Basel, University 
of Cambridge, Case Western Reserve University, University of Chicago, Drexel 
University, Fermilab, the Institute for Advanced Study, the Japan Participation 
Group, Johns Hopkins University, the Joint Institute for Nuclear Astrophysics, 
the Kavli Institute for Particle Astrophysics and Cosmology, the Korean 
Scientist Group, the Chinese Academy of Sciences (LAMOST), Los Alamos National L
aboratory, the Max-Planck-Institute for Astronomy (MPIA), the Max-Planck-Institute 
for Astrophysics (MPA), New Mexico State University, Ohio State University, 
University of Pittsburgh, University of Portsmouth, Princeton University, the 
United States Naval Observatory, and the University of Washington.


\end{document}